\documentclass[epj]{svjour}

\def\Journal#1#2#3#4{{#1} {\bf #2} (#3) #4}

\usepackage{graphics}
\begin{document}
\title{Beyond the quark model of hadrons from lattice QCD}
\author{Chris Michael  
}                     
\institute{Theoretical Physics Division, Mathematical Sciences Department, 
University of Liverpool, Liverpool L69 3BX, UK.}
\date{Received: date / Revised version: date}
%
\abstract{
 Lattice QCD can give direct information on OZI-violating contributions
to mesons. Here we explore the contributions  that split  flavour
singlet and non-singlet meson masses. I discuss in detail the spectrum 
and decays  for scalar mesons (ie including glueball effects). I also
review the status of hybrid mesons and their decays. 
\PACS{
      {12.38.Gc}{Lattice QCD calculations}  \and
      {12.39.Mk}{Glueball and non-standard multi-quark/gluon states}
     } 
} 
\maketitle
\section{Introduction}
\label{intro}

 Lattice QCD is a first-principles approach to solving QCD
non-perturbatively.  It enables theorists to explore the consequences of
QCD for quarks of different mass  and so complements experimental
studies. It is especially valuable for  validitating phenomenological
models. 
 Direct comparison with experiment is hampered by two obstructions: (i)
the continuum  limit (lattice spacing $a \to 0$) should be taken but this
is not computationally  feasible in many cases and (ii) the $u$ and $d$ quark
masses are too light to be  successfully explored on a lattice so that
one has to rely on chiral perturbation theory and extrapolate from heavier 
quark masses.
 Nevertheless, lattice QCD enables exploration of the hadron spectrum
and matrix elements  and can be used to quantify the success (and
failure) of quark models. 
 One area where the quark model is deficient is in flavour singlet
mesons. Here the  gluonic degrees of freedom are likely to make a
significant contribution.

\section{Flavour singlet mesons}
\label{sec:1}

 In lattice QCD, one studies mesons by creating a quark anti-quark pair 
at time 0 with the quantum numbers of that meson and then annihilating 
the quark anti-quark pair at time $t$.  The quark sources and sinks  are
then combined using the fully non-perturbative quark propagators 
determined on the lattice. For flavour singlet mesons, there will be two
 contributions: disconnected $D(t)$ and connected $C(t)$ as illustrated
in fig.~1.

\begin{figure}[h]

\begin{center}
\resizebox{0.2\textwidth}{!}{
  \includegraphics{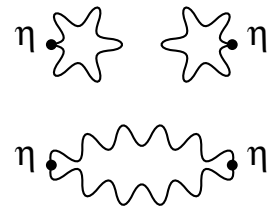}
 }
 \caption{ The  disconnected (above) and connected (below) correlators 
for flavour singlet mesons at time separation $t$. Here 
the wiggly lines represent full quark propagators. The illustration is for a 
pseudoscalar meson ($\eta$) but the same diagrams apply to all mesonic 
quantum numbers.
 }
 \label{dbyc}

  \end{center}
\end{figure}

At large t where ground state contributions dominate   these measured
correlations satisfy
$  C(t) =  c e^{-m_1 t}  $
 and 
 $  C(t)+D(t) = d e^{-m_0 t} $
 where  $m_0$ is the flavour singlet mass  and $m_1$ the flavour
non-singlet mass. 
 Then by a study of  $D/C$ which is given  by 
 $ (d/c) \exp((m_1-m_0) t) -1$
 one can explore the mass splitting between flavour singlet and
non-singlet which is a measure of the OZI violating gluonic contribution 
arising from the disconnected diagram.

 From a lattice study of this for all mesonic quantum
numbers~\cite{all-ozi} the only significant disconnected contributions
were found for scalar and for pseudoscalar mesons. This is not
unexpected: both from the phenomenology of the observed meson masses and
from the expectation that the lightest glueball  has  scalar quantum
numbers and that topological charge fluctuations have pseudoscalar 
quantum numbers.
 
 For a review of the situation concerning pseudoscalar mesons see
ref~\cite{eta}.
 We now discuss the glueballs and their impact on the meson spectrum.

 \section{Glueballs and scalar mesons}
\label{sec:2}

Glueballs are defined to be hadronic states made primarily from gluons.
The full non-perturbative gluonic interaction is included in quenched
QCD.  In the quenched approximation, there is no mixing between such
glueballs  and quark - antiquark mesons. A study of the glueball
spectrum in quenched QCD  is thus of great theoretical value. 

 This has been studied extensively~\cite{ukqcd,mpglue} and the consensus
is that the lightest glueball  has scalar quantum numbers with the
tensor ($J^{PC}=2^{++}$) and pseudoscalar  glueballs ($J^{PC}=0^{-+}$)
next in mass. The quenched results have been explored  to very small
lattice spacings and there is convincing evidence that the continuum
limit values have been extracted - see fig.~1. Since the quenched
approximation does {\em not} reproduce experiment,  different ways to
set the scale will differ - by $\pm 10\%$. Using a conventional  scale
assignment ($r_0 = 0.5 $ fm) gives masses of around 1.6 GeV for the
scalar  glueball and 2.2 GeV for the tensor.

One signal of great interest would be  a glueball with $J^{PC}$ not
allowed for $q \bar{q}$ - a spin-exotic glueball or {\em oddball} -
since it would  not mix with $q \bar{q}$ states. These states are
found~\cite{MTgl,ukqcd,mpglue} to be  high lying: considerably above
$2m(0^{++})$. Thus they are  likely to be in a region very difficult to
access unambiguously by experiment.

\begin{figure}[bt] 
\vspace{-2.5cm}  

\resizebox{0.49\textwidth}{!}{
  \includegraphics{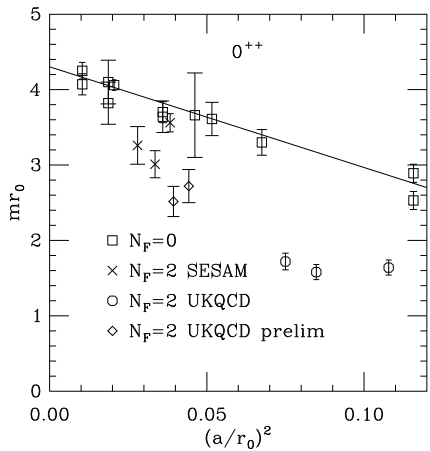}
 }
 \caption{ The value of mass of the  lightest $J^{PC}=0^{++}$  meson
(glueball) from quenched data
($N_F=0$){\protect\cite{DForc,MTgl,ukqcd,gf11}} in units of $r_0$ where
$r_0 \approx 0.5$ fm. The straight line  shows a
 fit describing the  approach to the continuum limit as $a \to 0$.
 Results~{\protect\cite{bali,cmcm176,cmcm202}} for the lightest
scalar meson with $N_F=2$ flavours of sea quarks are also shown.
   }
\end{figure}

 Within the quenched approximation, the glueball states are unmixed 
with $q \bar{q},\ q \bar{q} q \bar{q}$, etc. Furthermore, the  $q
\bar{q}$ states have degenerate flavour singlet and non-singlet states
in the quenched approximation.  This gives rise to  anomalies in the
quenched approximation:  for example the scalar meson propagation can
have the wrong sign~\cite{fnal} because the $\eta \pi$  intermediate state is
mistreated.  Once quark loops are allowed in the vacuum then these
anomalies are removed. Indeed, for the favour-singlet states of any
given $J^{PC}$,  there will be mixing between the  $s \bar{s}$ state,
the  $u \bar{u}+d \bar{d}$ state  and the glueball. 
 Since the scalar glueball is lightest, we expect the largest effects
here and one can  explore this by measuring directly the scalar  mass
eigenstates in a study with $N_f=2$ flavours of sea-quark.

 Most studies have shown no significant change of the glueball spectrum
as dynamical quarks are included~\cite{bali}.  However the
larger lattice spacing result~\cite{cmcm176} shows a significant 
reduction in the lightest scalar mass, as shown in fig.~1.
 Before concluding that this implies a lower scalar mass in  the
continuum limit, one needs to check whether  an enhanced order $a^2$ 
correction might be present. Studies using the same approach  at a finer
lattice spacing~\cite{ht,cmcm202} do suggest that this  large order $a^2$
effect is significant, but studies even nearer to the continuum  or with
improved actions are needed to resolve this fully.

Let us now discuss the mixing of the scalar glueball and scalar mesons.
In quenched  QCD  there is the problem described above concerning
scalar mesons. For heavy enough quarks this is unimportant and one 
can measure the mixing strength on a quenched lattice even
though no mixing actually occurs. On a rather coarse lattice ($a^{-1} 
\approx 1.2$ GeV), two groups  have attempted this~\cite{weinssg,cmcm176}.
Their results expressed as the mixing for two degenerate quarks of mass
around the strange quark mass  are similar, namely $E \approx 0.36$
GeV~\cite{weinssg} and 0.44 GeV~\cite{cmcm176}.
 Opinions differ~\cite{weinssg,cmcm176} as to whether this large mixing
which would shift the glueball mass down by 20\% will persist to the
continuum  limit.

 From dynamical fermion studies with $N_f=2$, one can determine the 
flavour singlet and non-singlet mass spectrum. No glueball, as such, 
can be defined. What we find~\cite{cmcm176,cmcm202} is that the lightest
flavour-singlet scalar  meson is lighter than the lightest flavour
non-singlet. This is in  qualitative agreement with the mixing scenario
described above.  

As well as this mixing of the glueball with $q \bar{q}$ states, there
will be  mixing  with $q \bar{q} q \bar{q}$ states which will be
responsible for the  hadronic decays. A first attempt to study
this~\cite{gdecay} in quenched QCD yields an estimated width for decay
to two pseudoscalar mesons from the scalar glueball of order 100 MeV.  A
more realistic study  would involve taking account of mixing using 
gluonic, $q \bar{q}$ and $q \bar{q} q \bar{q}$ operators  in full QCD.

\section{Hybrid mesons}
 \label{sect:3}

 A hybrid meson has the gluonic degrees of freedom which are  excited
non-trivially. The most significant consequence of this, experimentally,
 will be mesons with $J^{PC}$ values not allowed in the quark model for
a $q \bar{q}$ system (such as $1^{-+}$). 

 On the lattice one can easily deal with relatively light quarks (down
to the strange quark mass)  or extremely heavy quarks (treated as static
or using NRQCD). 
 I first discuss hybrid mesons with static heavy quarks where the
description  can be thought of as an excited colour string. The lattice
results~\cite{liv,pm,jkm} can then be presented as potential energy
versus  quark-antiquark separation $R$. The ground state will correspond
to the  usual interquark potential whereas excited states will have
non-trivial  representations of the symmetries, e.g. the lightest
excited state ($\Pi_u$) has colour flux which is a   difference of paths
from quark to antiquark of the form $\, \sqcap - \sqcup$.

 From the potential corresponding to these excited gluonic states, one
can  determine the spectrum of hybrid quarkonia using the Schr\"odinger
equation in the Born-Oppenheimer approximation. 
  The $\Pi_u$ symmetry state will produce a degenerate set of eight 
hybrid mesons of which those with   $J^{PC}=  1^{-+},\ 0^{+-}$ and  
$2^{+-}$  are spin-exotic and hence will not mix with $Q\bar{Q}$ states.
They thus form a very attractive goal for experimental searches for
hybrid  mesons.

 Within the quenched approximation,  the lattice evidence  for
$b\bar{b}$ quarks points to a  lightest hybrid spin exotic meson H with
$J^{PC}=1^{-+}$ at an energy given by~\cite{hf8}   $(m_H-m_{2S})r_0=1.9
\pm 0.1$. This has been checked  with $N_f=2$
studies~\cite{cppacs,hdecay}  which find similar mass ratios
$(m_H-m_{1S})/(m_{1P}-m_{1S})$ but a larger splitting
in terms of $r_0$,  namely~\cite{cppacs} $(m_H-m_{2S})r_0=2.4 \pm 0.2$. 
 Using the experimental mass of the $\Upsilon(2S)$, these results imply
that the lightest spin exotic  hybrid is in the mass range 10.7 to 10.9
GeV. Above this energy there will be many more hybrid  states, many of
which will be spin exotic.

 Within this static quark framework, one can explore the decay
mechanisms.  One special feature is that the symmetries of the quark and
colour fields about the static quarks must be preserved exactly in
decay.  This has the consequence that the decay from a $\Pi_u$ hybrid
state to the open-$b$ mesons ($B \bar{B},\   B^* \bar{B},\  B
\bar{B^*},\ B^* \bar{B^*}$) will be forbidden~\cite{hdecay} if the 
light quarks in the $B$ and $B^*$ mesons are in an S-wave relative to
the heavy quark. The decay to $B^{**}$-mesons with light quarks in a
P-wave is allowed by symmetry but not energetically. 
 The only allowed decays are to $\chi_b +M$ where $M$ is  a light
quark-antiquark meson in a flavour singlet. Lattice
estimates~\cite{hdecay} of these  transitions have been made and the
dominant mode (with a width of around 100 MeV) is found to  be with $M$
as a scalar meson. 
 This decay analysis does not take into account heavy quark motion or
spin-flip  and these effects will be significantly more important for
charm quarks than for $b$-quarks. 

Several lattice groups~\cite{livhyb,milc,sesamhyb,ml} have studied
hybrid spectra for light quarks  and  find the lightest spin-exotic
hybrid to have $J^{PC}=1^{-+}$ and mass  (for $s$-quarks) of around
2GeV. The corresponding light-quark ($u$, $d$) state would be  around
120 MeV lighter. The light quark results have also been~\cite{milc}
extrapolated to charm quarks and masses near 4.4GeV  are found for the
corresponding state. A recent study has confirmed this value but with a
very long extrapolation~\cite{ml}. These mass estimates can be compared
to naive estimates~\cite{pm} of  the spin-exotic charm  state mass of
4.0 GeV from the static quark approach  which will have an uncontrolled
systematic error. 
 
It is not easy to reconcile these lattice results  with experimental
indications~\cite{expt} for resonances at 1.4 GeV and 1.6 GeV,
especially the  lower mass value.  Mixing  with  $q \bar{q} q \bar{q}$
states such as $\eta \pi$ is not included for realistic quark masses in
the  lattice calculations. This can be interpreted, dependent on one's
viewpoint,  as either that the lattice calculations  are incomplete or
as an indication that the experimental states may have an  important
meson-meson component in them.

 An attractive prospect to study hybrid mesons is from $p \bar{p}$
annihilation.  Spin-exotic hybrids, which provide the best signal, will
only be produced in association  with other hadrons. The decay channel to 
$\chi_c + \pi +\pi$ with the two $\pi$ mesons in an S-wave is a promising 
detection channel. Lattice studies give first estimates of the masses 
and decay widths but more work needs to be done to constrain  the systematic 
errors on these estimates.


\end{document}